# Machine learning driven search of hydrogen storage materials

T. Banerjee,[1,2,+] K. Ji,[1,+] W. Xia,[1] G. Ouyang,[1] T. Del Rose,[1] I. Z. Hlova,[1] B. Ueland,[1] D. D. Johnson,[1,3]

C.-Z. Wang,[1] G. Balasubramanian,[2] P. Singh[1,*]

[1] Ames National Laboratory, US Department of Energy, Iowa State University, Ames, IA 50011, USA
[2] Department of Mechanical and Industrial Engineering, University of New Haven, West Haven, CT, 06516, USA
[3] Department of Materials Science & Engineering, Iowa State University, Ames, IA 50011, USA

**Abstract**

The transition to a low-carbon economy demands efficient and sustainable energy-storage solutions, with hydrogen emerging as a promising clean-energy carrier and with metal hydrides recognized for their hydrogen-storage capacity. Here, we leverage machine learning (ML) to predict hydrogen-to-metal (H/M) ratios and solution energy by incorporating thermodynamic parameters and local lattice distortion (LLD) as key features. Our best-performing ML model provides improvements to H/M ratios and solution energies over a broad class of ternary alloys (easily extendable to multi-principal-element alloys), such as Ti-Nb-X (X = Mo, Cr, Hf, Ta, V, Zr) and Co-Ni-X (X = Al, Mg, V). Ti-Nb-Mo alloys reveal compositional effects in H-storage behavior, in particular Ti, Nb, and V enhance H-storage capacity, while Mo reduces H/M and hydrogen weight percent by 40-50%. We attributed to slow hydrogen kinetics in molybdenum rich alloys, which is validated by our pressure-composition isotherm (PCT) experiments on pure Ti and $Ti_5Mo_{95}$ alloys. Density functional theory (DFT) and molecular simulations also confirm that Ti and Nb promote H diffusion, whereas Mo hinders it, highlighting the interplay between electronic structure, lattice distortions, and hydrogen uptake. Notably, our Gradient Boosting Regression model identifies LLD as a critical factor in H/M predictions. To aid material selection, we present two periodic tables illustrating elemental effects on (a) $H_2$ wt% and (b) solution energy, derived from ML, and provide a reference for identifying alloying elements that enhance hydrogen solubility and storage.

***Keywords***: *Alloys, ML, DFT, MD, Hydrogen storage, Solution energy*

[+] Equal contribution
[*] Corresponding author: psingh84@ameslab.gov/prashant40179@gmail.com

## 1. Introduction

Transitioning to a low-carbon economy, hydrogen (H) increasingly garners attention as a key clean-energy carrier in the shift toward renewable, efficient, and sustainable energy source. It can be produced through electrolysis, using excess electricity from renewable sources and this stored hydrogen can then be utilized in various ways to generate electricity during periods of high demand or, when renewable generation is low, providing a means to balance these intermittent energy sources [1–3]. Hydrogen has high energy density by weight making it an attractive option for storing and transporting energy efficiently [4]. However, maximizing hydrogen's potential as an energy carrier faces continued hurdles, especially in storage and transportation. Metal hydrides are attractive H-storage materials because of their high storage capacities and favorable properties [5,6].

Complex solid-solution alloys (CSA) or multi-principal-element alloys (MPEA) are one such class of materials with vast numbers of unexplored compositions for adjusting desired properties, like H-storage behavior. However, large CSA design space limits efficient exploration using Edisonian "trial-and-error" methods in experiments and computations [7–12]. Artificial-intelligence (AI)/ machine-learning (ML) methods, aided by extensive databases of material structures and properties, has extended design space by enabling robust high-throughput predictions of material properties. At the same time, it reveals correlations and patterns between physical factors that conventional methods might overlook.

AI/ML applications to predict improved metal hydrides for hydrogen storage is a complex and growing area of research. Several studies have focused on developing AI/ML platforms to enhance the property prediction for increasing hydrogen storage and its efficiency [13–16]. Consecutive regression and multi-class classification algorithms were used to predict H-weight percentage (wt%) and classify the hydride class, the key parameters for assessing H-storage capacity of metal hydrides [13,17]. By leveraging ML, its aim is to categorize metal hydrides via their characteristics and properties, yielding insights into identifying more efficient H-storage materials [17]. Furthermore, utilizing hybrid-ML framework using deep multi-layer perceptron (MLP) in identifying best-in-class H-storage metal hydrides demonstrates effectiveness of combining different ML approaches to enhance predictive accuracy [14] by overcoming associated high dimensionality of data and challenges imposed by the scarcity of experimental data. Literatures show that alloy structure and compositions along with several thermodynamic parameters are key to achieving higher H-storage capacity [9,15,16,18,19]. Leveraging explainable ML, Witman, *et al.* [15] showed that volume-based descriptor (calculated per atom) has a significantly strong correlation with equilibrium pressure that determine the tendency to form metal hydrides and their stability, whereas Lu *et al.* [19] studied the structure-property relationship of hydrides using ML and reported that the lattice constant, valency-electron concentration (VEC), and $Z/r^3$ (where Z and r are the valence electron number and atom radius) play key roles in the H-storage capacity of quaternary Ti-Cr-Fe-V alloy.

Molecular simulations are widely used to study H-storage properties to provide insights into H-adsorption capacities and binding energies, kinetics of H diffusion/adsorption, reaction pathways for H adsorption/dissociation, formation in H-evolution reactions [20–24], and on other material properties. Additionally, density functional theory (DFT) informed interpretable ML using Bayesian optimization shows that body-centered cubic (BCC) multi-principal elemental alloys (MPEA) theoretically exhibit high hydrogen storage (2.83 wt%), and associated mechanical strength and electronic configurations [25]. Furthermore, BCC alloys exhibit a H/M=2+ by tuning alloy composition, and H-storage properties can be modified [26]. Suwarno, *et al.* [16] applied ML models to show the effects of alloy elements on H-storage capacity and solution energy, giving insight into alloy selection of elements for experiments. Other than metallic alloys, ML-based studies have been efficient in characterizing H-storage capacities of non-metallic materials, including metal-organic frameworks (MOFs) [27,28], $LiBH_4$-based [29], and porous carbon-based materials [30]. Results show that several functional groups for MOFs and other thermodynamic parameters like the temperature, pressure and surface properties (like pore volume and surface pore fractions) play a major role in hydrogen uptake [11].  Overall, these studies reflect the significant role that ML can play by harnessing the combined power of computational methods of MD, DFT, and data-driven ML algorithms, to accelerate the discovery and optimization of metal hydrides with enhanced H-storage.

Here, we present a ML-guided approach that integrates fundamental ML principles, such as feature selection, model training, and validation, with atomistic simulations for the predictive design of solid-state H-storage materials. We utilize the US Department of Energy (DOE) H-storage dataset, which includes features like H wt.%, solution energy, entropy of formation, pressure, temperature, and other thermodynamic parameters across various chemical compositions. A key addition to our model is the inclusion of local lattice distortion (LLD), resulting from atomic-size mismatch between elements, shown to correlate with H-storage performance in MPEAs [25]. As a distinct difference between our model and others, we emphasize that H/M is more informative than wt.%, as it is a more effective descriptor for alloy structure and H-storage. We show that AI/ML framework can facilitate high-throughput search of H-storage materials over multi-dimensional alloy design space, exemplified for ternary metal hydrides with BCC and FCC crystal-structure, i.e., Ti-Nb-X (X=Mo, Cr, Hf, Ta, V, Zr) and Co-Ni-X (X=Al, Mg, V).

We also performed DFT and MD simulations to reveal the role of hydrogen-metal interaction on electronic-structure, local-lattice distortion (LLD), and H diffusion in affecting H-storage capacity on selected Ti-Nb-Mo and Co-Ni-X alloys with best H/M and H solubility. The correlation between H-storage capacity and solution energy with compositional wt.% has been shown, primarily focusing on refractory elements like Ti, Nb, and Mo. Overall, our study underscores the significant role that combining DFT, MD, and data-driven AI/ML algorithms can play in accelerating discovery and optimization of metal-hydrides with enhanced H-storage properties.

## 2. Methods

**2.1 Data selection, preparation and feature engineering: In Fig. 1a, w**e have shown the hydrogen storage (HyStr) experimental database consisting of $H_2$wt% and hydrogen solution energy of nearly 1700 metal alloys as a function of absorption temperature. The experimentally recoded database contains diverse classes of metal alloys including $A_2B$, AB, $AB_2$, $AB_5$, MIC, SS, Mg, Low/Medium/High-entropy Alloys.

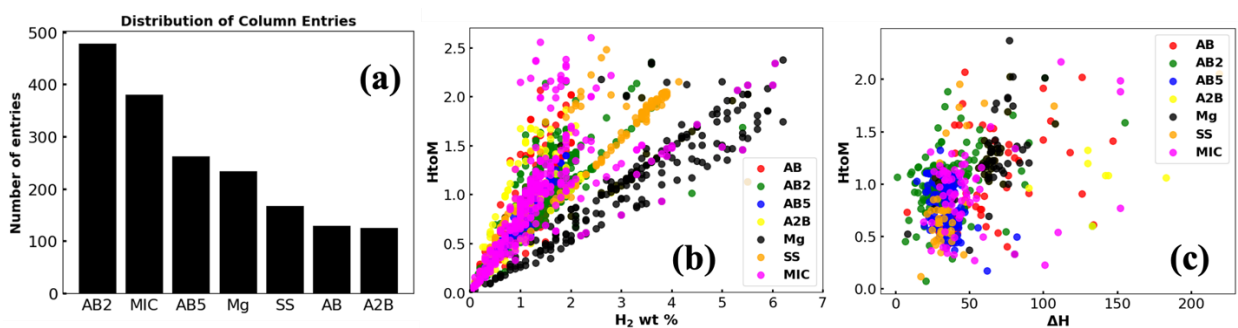

**Figure 1. (a)** Distribution of different types of compounds present in the data after data preparation and outlier removal. **(b)** The Spearman correlation between H/M with H wt.% shows a strong positive linear dependence with a Spearman correlation factor of 0.85, whereas **(c)** H/M and Solution energy are quite uncorrelated. This suggests that H/M and Solution energy models are independent. All different classes are plotted in different colors.

We have shown the dependency of H/M with H wt.% in **Figure 1(b)** and from this, we clearly see a strong positive linear correlation (Spearman correlation factor of 0.85) between these two. H/M shows the atomic ratio of hydrogen atoms relative to the other atoms of the compounds that measure the concentration of hydrogen atoms whereas H wt.% in a material is calculated as the mass of hydrogen divided by the total mass of the material, calculated in percentage. Both measures provide insights into the hydrogen content of alloys, but they quantify the presence of hydrogen in different ways and the positive correlation in **Figure 1(b)** physically justifies that well. **Figure 1(c)** shows how two of our predictors H/M and solution energy for hydrogen storage are correlated and we see that H/M and solution energy as features are mostly independent of each other, showing a random unbiased data distribution. **Figure 1(b)** also shows that Mg-based compounds has a wide range of H2 wt% capacity of ~5-6 wt% whereas the AB2 and A2B types can only store ~2-3 wt% of hydrogen which is critical for further analysis shown in this study.

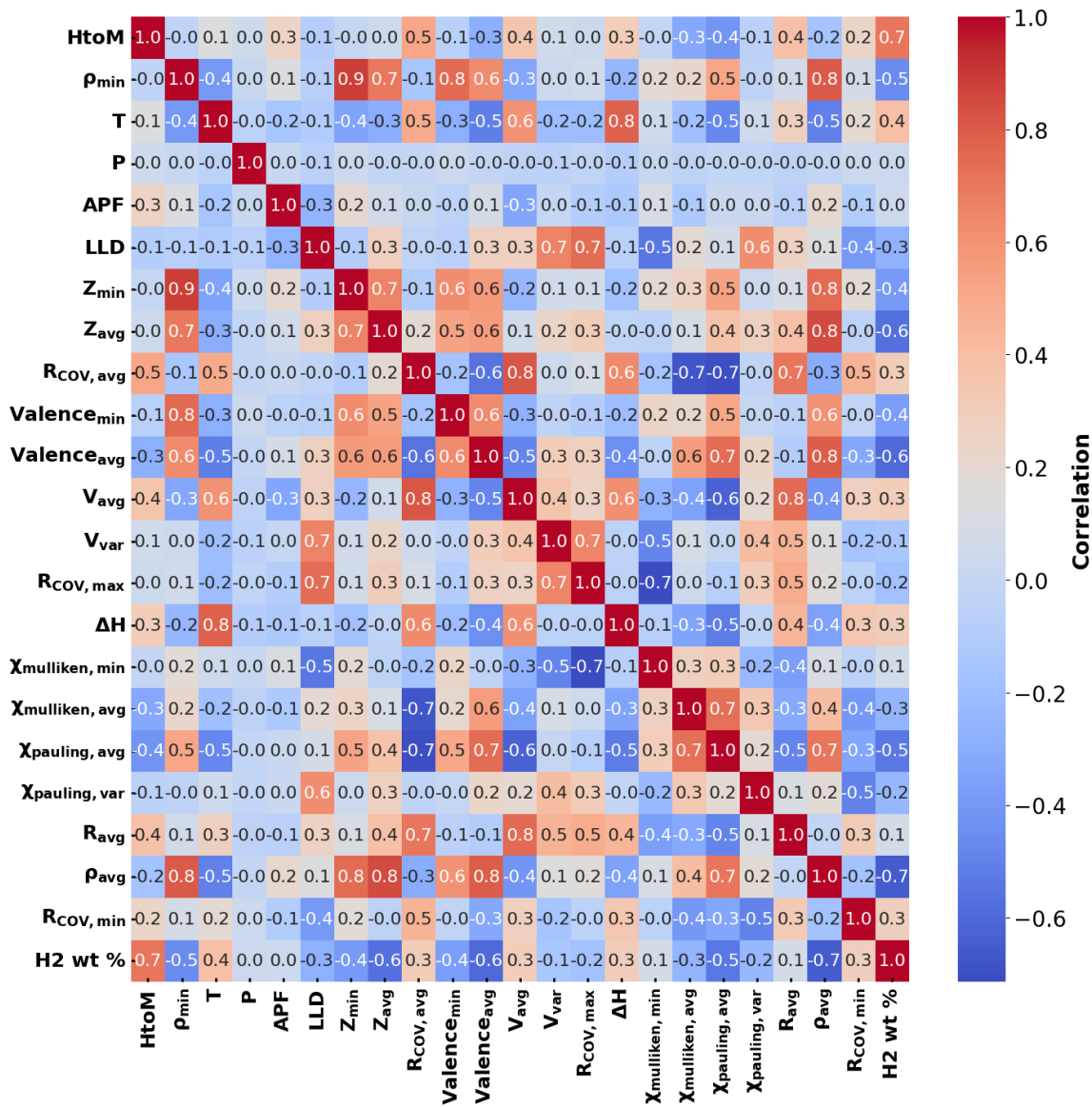

**Figure 2.** The Pearson correlation matrix shows the correlations between all the features with each other present on the dataset for the ML model training. It can be seen clearly that H/M has a strong positive correlation (+0.7) with H wt.% whereas Solution Energy is strongly correlated with Temperature and Entropy of formation, which is physically significant.

The Pearson correlation matrix for all the features combined is shown in **Figure 2**, indicating the correlation measures (Fig. 2a) the linear relationship between two features and assumes that the variables are normally distributed. Likewise, the Spearman correlation coefficient (**Fig. 2b**) assesses the strength and direction of a monotonic relationship, making it more robust to non-linear associations and outliers, regardless of the linearity of the relationship [32]. The Pearson correlation between H/M to solution energy is 0.3 and H/M to hydrogen wt% is 0.7. **Figure 2** shows that covalence ($R_{COV}$), average atomic radius ($R_{avg}$), average atomic volume ($V_{avg}$), and atomic packing fractions (APF) show strong positive correlations with both the H/M and solution energy of metal hydrides, indicating the importance of these atomic features to improve the accuracy of the ML model. Also, we see that LLD does not show any significant correlation with any of the thermodynamic parameters, and it has a -0.3 correlation with hydrogen wt.%, however, as LLDs are calculated based on the atomic mismatch between constituent elements, it shows a strong correlation (+0.7) with covalent radius ($R_{cov}$).

**2.2 Feature engineering:** Initially, for the complete dataset selected for the study, only the thermodynamic features like pressure, temperature, solution energy are there alongside each composition. Because of this scarcity of features, we have added different statistical-based atomic and electronic features like, Pauling electronegativity ($\varkappa_{pauling}$), Mulliken electronegativity ($\varkappa_{mulliken}$), covalent, atomic radii, atomic volume, weight, density, LLD, etc. and we approached our model development which could lead to attaining relatively higher accuracy without bias and overdependence on specific features. Mostly, we added the mean, variance, minimum and maximum values of each of those atomic, electronic and material features and trained the model using only these feature sets. The mean and variance were both weighted according to the proportion of the element within each alloy structures. Mathematically, the mean (μ), variance ($\sigma^2$), minimum (min), and maximum (max) values of a feature $X = \{x_1, x_2, ...., x_n\}$ are calculated as, $\mu = \frac{1}{n}\sum_{i=1}^{n} x_i$; $\sigma^2 = \frac{1}{n}\sum_{i=1}^{n}(x_i - \mu)^2$; $\min(X) = \min\{x_1, x_2, x_3, ..., x_n\}$; and $\max(X) = \max\{x_1, x_2, x_3, ..., x_n\}$. Additionally, atomic packing fraction (APF), a measure of how densely the atoms is closely packed in each structure, is added as a feature which is calculated by the fraction of volume in a crystal structure that is occupied by the constituent atoms. APF can be a critical feature in hydrogen storage for alloys, as it is strongly correlated with the availability of interstitial sites where hydrogen atoms can reside within lattice structures. The APF is calculated by, $APF = \frac{\frac{4}{3}\pi \rho N_a \sum(r_i^3 c_i)}{\sum M_i c_i}$, where $\rho$ is the density of the material, $r_i$, $c_i$, and $M_i$ are the atomic radii, compositional fraction, and molar mass of i-th element respectively, and $N_a$ is Avogadro's number. By addition of all these atomic, electronic, and structural features; calculated from elemental properties using the alloy compositions of the dataset; we finally reached a total of 38 features for each composition. This statistical approach helps in reducing bias and prevents over-reliance on all thermodynamic features, thus improving the model's accuracy without overfitting, which is the primary objective of this work.

All compositions that could not be directly parsed as compositions, such as "$Mg_2Ni_{1-y}Be_y$" were removed. In addition, any values for temperature, pressure and heat of formation that were outliers by more than 5 standard deviations were removed, as well as values that were non-numerical or incorrectly formatted. After pre-processing, our sample had 1775 compositions and of these, only 1100 compositions had temperature and/or pressure values, and only 558 had values for heat of formation/solution energy. Furthermore, we used the k nearest neighbor (KNN) type supervised learning technique which has been effective in predicting missing values of a dataset to make predictions based on the mean of the 5 nearest data points. KNN-type supervised learning method computes the average or weighted average of their target values which is used as the predicted value for the new data points [33,34].

Evaluating the performance of machine learning models is a critical step in assessing their effectiveness. Here we have considered 4 parameters- R-squared ($R^2$) score, Mean Absolute Error (MAE), Root Mean Squared Error (RMSE), and Mean Absolute Percentage Error (MAPE), for evaluating our models' performance (**Eq. 1-4**). The R-squared score, also known as the coefficient of determination, first introduced by Wright [35] measures the proportion of the variance in the target variable that is explained by the model i.e., it shows how well the dependent variable is evaluated by all the independent variables [36]. It varies from 0 to 1 where a higher $R^2$ score means a better fit of the dataset. MAE is straightforward and measures the average absolute deviation between the model's predicted values and the actual target values. RMSE penalizes heavily to outliers thus making it more sensitive to outliers in a dataset, whereas MAPE focuses on percentage error and becomes effective in quantifying the relative variations of the predicted data and actual data, but it is ineffective when large errors are determined [36].

$$R2 = 1 - \frac{\sum(y_i - \hat{y})^2}{\sum(y_i - \bar{y})^2} \tag{1}$$

$$MAE = \frac{1}{N}\sum_{i=1}^{N}|y_i - \hat{y}| \tag{2}$$

$$RMSE = \sqrt{\frac{1}{N}\sum_{i=1}^{N}(y_i - \hat{y})^2} \tag{3}$$

$$MAPE = \frac{1}{N}\sum_{i=1}^{N}\frac{|y_i - \hat{y}|}{|y_i|} \times 100\% \tag{4}$$

where $\bar{y}$ is the mean value of y, $\hat{y}$ is the predicted value of y, and $y_i$ is the i-th data point.

Although it is almost impossible to predict the performance of a machine-learning regression task just by one single evaluation parameter, studies show that the $R^2$ score (i.e., the coefficient of determination) becomes the most informative and truthful than any other performance matrices like MAE, MAPE or RMSE [36].

Additionally, the atomic size mismatch adopted to measure the LLD effect here is calculated as [35],

$$\delta = 100 \sqrt{\sum_{i=1}^{n} c_i \left(1 - \frac{r_i}{\bar{r}}\right)^2} \tag{5}$$

Where $c_i$ and $r_i$ are the atomic percentage and radius of individual constituent elements of the alloy respectively.

**2.3 Training of ML Models:** Due to the scarcity of data points, we have applied several classical machine learning algorithms, mainly linear, decision tree, and ensemble-based models, and performed necessary model tunings as per hyperparameter optimization to better the accuracy. First, we started by applying the classical ML models like Linear regression, Lasso, and Bayesian Ridge regressions and later upgraded to some ensemble and tree-based models like Gradient Boosting Regressor (GBR), Extreme Gradient Boosting Regressor (XGB), and Random Forests (RF). Linear regression, the simplest ML algorithm, assumes all the features contribute equally to the outcome and it is well-suited to capture linear relationships by minimizing the least squared differences whereas Lasso (Least Absolute Shrinkage and Selection Operator) and Ridge regressions introduce L1 and L2 regularizations respectively by adding a penalty term based on the absolute values of the coefficients (L1) or the sum of squared coefficients (L2) which can improve generalization by preventing overfitting. On the other hand, GBR, XGB, and RFs are powerful ensemble learning techniques and among them, GBR is an ensemble technique that builds a strong predictive model by combining the predictions of multiple weak learners, typically decision trees sequentially, and XGB is an optimized implementation of GBR designed for better accuracy and performance that extends traditional GBR by adding regularization (L1 and L2 both) terms in its objective function to control model complexity and reduce overfitting. RF model constructs a multitude of decision trees during training and outputs the mean prediction of the individual trees where each tree is trained on a random subset of the training data and features, which also helps reduce overfitting and improve generalization.

The performance of machine learning models greatly depends on the hyperparameters before training the models. Hyperparameters are the configuration settings that are external to the model itself and cannot be learned from the training data and the optimization of these parameters becomes important as it directly impacts the generalization, efficiency, and stability of the model [37,38]. Hyperparameter values significantly affect the performance of an ML model whereas poorly chosen hyperparameters can result in longer training times, increased bias, and overfitting. Various machine learning algorithms come with distinct sets of hyperparameters that profoundly influence their performance. Hyperparameter tuning is essential for identifying the most favorable combination of these values, ultimately leading to the highest model performance [39]. For this purpose, we employed the "GridSearchCV" from Scikit-learn [40] which facilitates an exhaustive search through a predefined randomized hyperparameter grid across a specified range. The models subjected to hyperparameter optimization include ensemble models like GBR, RF, and XGB, as well as Lasso and Ridge regression.

After conducting hyperparameter optimization using GridSearchCV, we identified the best-performing hyperparameters for each of the best 3 predictive models. The selection criteria were based on the model's performance metrics, such as accuracy, precision, recall, and mean squared error averaged over 5-fold cross-validation. **Table 1** the top three hyperparameters for each model that yielded the best results:

**Table 1.** Hyperparameters (HP) for best models after optimization via exhaustive Gridsearch algorithm for the 3 predictors analyzed in this study.

| Models | Best model | HP1 | HP2 | HP3 | HP4 |
|---|---|---|---|---|---|
| H/M | Random Forest | max_features=4 | max_depth=14 | n_estimator=500 | min_samples_split=6 |
| Solution Energy | Extreme Gradient Boost | learning_rate=0.08 | max_depth=4 | n_estimator=300 | N/A |
| H$_2$ wt % | Gradient Boost | learning_rate=0.05 | max_depth=4 | n_estimator=1500 | min_samples_split=10 |

These hyperparameters were obtained as they led to improved performance of the models making the prediction more efficient. By fine-tuning these parameters, we were able to enhance the generalization capability of the models, reduce training times, and mitigate the risks of overfitting, thereby achieving a more robust performance.

**2.4 MD simulation of H diffusion:** The interatomic potentials for the molecular simulations in this study can be represented as, $U = U_{M-M} + U_{H-H} + U_{M-H}$, where $U_{M-M}$ denotes the metal-metal interaction, and $U_{H-H}$ and $U_{M-H}$ represent the hydrogen-hydrogen and metal atom-hydrogen interactions respectively. Lennard Jones (LJ) 12-6 potential, presented in equation (6) [41,42] was applied with a cut-off distance of 10 Å to model the van der Waals interactions between each of these metal elements with gaseous hydrogen atoms ($U_{M-H}$ and $U_{H-H}$).

$$U_{LJ}(r_{ij}) = 4\in_{ij}\left[\left(\frac{\sigma_{ij}}{r_{ij}}\right)^{12} - \left(\frac{\sigma_{ij}}{r_{ij}}\right)^{6}\right], r_c < r_{ij} \qquad (6)$$

Where, $\sigma_{ij}$, $\epsilon_{ij}$ were the distance where potential energy becomes zero and the potential wall depth respectively, and $r_{ij}$ is the distance between one LJ site to another. Lorentz-Berthelot mixing rule was applied to model the interaction parameters ($\sigma$ and $\epsilon$) for each of the $U_{H-H}\ and\ U_{M-H}$ cross-interactions [41]. Inter and intra-atomic LJ interaction parameters for TiNbMo combinations with atomic hydrogen were taken from reported literature [43,44].

Each simulation supercell initially consisted of 60 atoms depending on the different elemental compositions selected, and we have added 6 hydrogen atoms at the octahedral positions within the lattice structure to assess the hydrogen diffusion inside bulk metallic structures. First, the energy minimization of each structure was performed using the conjugate gradient method [44], followed by isothermal-isobaric (NPT) equilibration at 300 K, and constant pressure heating to 1000 K at 4.7 K/ps and a subsequent NPT equilibration at 1000 K to relax the structures. Finally, the hydrogen diffusion simulations were carried out on the relaxed structures at 1000 K. The average mean squared displacements (MSD) and subsequent radial distribution functions (RDF) were generated to qualitatively analyze hydrogen diffusion in different binary and ternary TiNbMo-based alloy systems. All the simulations were carried out using the Large-scale Atomic/Molecular Massively Parallel Simulator (LAMMPS) software [45] and a 1 femtosecond (fs) timestep size was used with velocity verlet algorithm [46,47] for performing the time integration of the equation of motions and the system temperature was controlled using Nosé-hoover thermostat [44,48,49].

**2.5 DFT methods**: We employed first-principles DFT as implemented in the Vienna Ab-initio Simulation Package (VASP) [50,51] for geometrical optimization (e.g., lattice constants, volume, bond length, and bond angles) and charge self-consistency. The generalized gradient approximation of Perdew, Burke, and Ernzerhof (PBE) was employed in all calculations [52] with a plane-wave cut-off energy of 520 eV. The choice of PBE over LDA or meta-GGA [53,54] functionals is based on the work of Söderling et al. [55] and Giese et al. [56] that establishes the effectiveness of GGA functionals. Large Supercell Random Approximates (SCRAPs), i.e., 60 atoms per cell, with the optimized disorder (zero-correlation) were generated (a single, optimized configurational representation) for DFT calculations. The energy and force convergence criterion of $10^{-8}$ eV and $10^{-6}$ eV/Å, respectively, were used for full (volume and atomic) relaxation of SCRAPs. The Monkhorst-Pack *k*-mesh was used for Brillouin zone integration during structural optimization and charge self-consistency calculations [57].

**3. Results and discussions**

By incorporating H/M, solution energy, and H2 weight percent as predictive metrics, we discuss below that how ML helps filter out less promising materials early, reducing the need for extensive and computationally expensive DFT validation. This targeted approach was not only used to save time and resources but also increases the likelihood of discovering MPEAs with optimized hydrogen storage performance, driving faster progress in the field.

**3.1 H/M, Solution Energy and $H_2$ wt.% predictive models**

The H-storage capacity in metals and alloys can be defined generally in 2 distinctive ways though they are closely related, i.e., (i) $H_2$ wt.% and (ii) H/M ratio. Typically, the gravimetric H-storage capacity is calculated from H wt.% which is the ratio of the mass of hydrogen absorbed within the compound to the mass of the compound including the stored hydrogen [58], which is given as,

$$H_2\ wt\% = \left(\frac{(H/M)M_H}{M_C + (H/M)M_H} \times 100\right)\%$$

where $M_H$ and $M_C$ are the molar masses of hydrogen and the material/compound storing hydrogen, and H/M is the atomic ratio of hydrogen to the H-storage material.

Here, we have developed a data-driven model to predict the H/M ratio. **Table 2** shows that RF performs the best in predicting H/M with a test set $R^2$ score of 74% with only the atomic, electronic and crystal properties of the alloy compositions. Additionally, the model was subjected to 5-fold cross-validation that predicts the robustness and generalizability, resulting in a mean $R^2$ score of 71%. Despite the relatively high $R^2$ score, the MAPE was >40% suggesting a substantial deviation of some instances from the model's prediction. This discrepancy highlights that while the model captures the overall trend in the data, it may not be accurate in predicting all individual instances. The GBR model demonstrated strong predictive capabilities for $H_2$ wt%, achieving a test set $R^2$ of 0.85 and a cross-validation $R^2$ score of 82% which is substantially higher than the H/M model showcasing the features used for predicting $H_2$ wt% were able to capture the underlying patterns and able to effectively generalize the target variable whereas H/M prediction becomes inherently more complex or noisy.

**Table 2.** Different evaluation parameter values ($R^2$, Mean Absolute Error (MAE), Root Mean Squared Error (RMSE), and Mean Absolute Percentage Error (MAPE)) for the H/M, $H_2$ wt%, and solution energy models.

| Models | Best performing ML model | Training data | | | | | Test data | | | | |
|---|---|---|---|---|---|---|---|---|---|---|---|
| | | $R^2$ | MAE | RMSE | MAPE | CV | $R^2$ | MAE | RMSE | MAPE | CV |
| H/M | RandomForest | 0.89 | 0.11 | 0.16 | 22.51 | 0.70 | 0.74 | 0.17 | 0.24 | 41.60 | 0.71 |
| Solution Energy | Extreme GradientBoost | 0.99 | 1.05 | 1.62 | 3.75 | 0.66 | 0.74 | 9.82 | 17.35 | 22.18 | 0.75 |
| $H_2$ wt. % | GradientBoost | 0.98 | 0.06 | 0.17 | 7.26 | 0.77 | 0.85 | 0.26 | 0.44 | 33.6 | 0.82 |

**Figure 3a-b** shows the training and test data parity plot for the H/M model and **Figure 3c** shows the top 10 important features dominating the H/M prediction using the RF model. Average radius, covalent

radius, and average electronegativity are the top 3 critical parameters, and interestingly, atomic packing fraction (APF), another important lattice parameter denoting the volumetric occupancy of atoms in the unit cell becomes a physically significant feature because hydrogen's susceptibility to adsorb in the interstitial gaps of lattice. The predictive model for $H_2$ wt% also shows similar important feature lists, however, average molecular weight ($Z_{avg}$) becomes the most important feature based on the GBR model (**Figure 3i**).

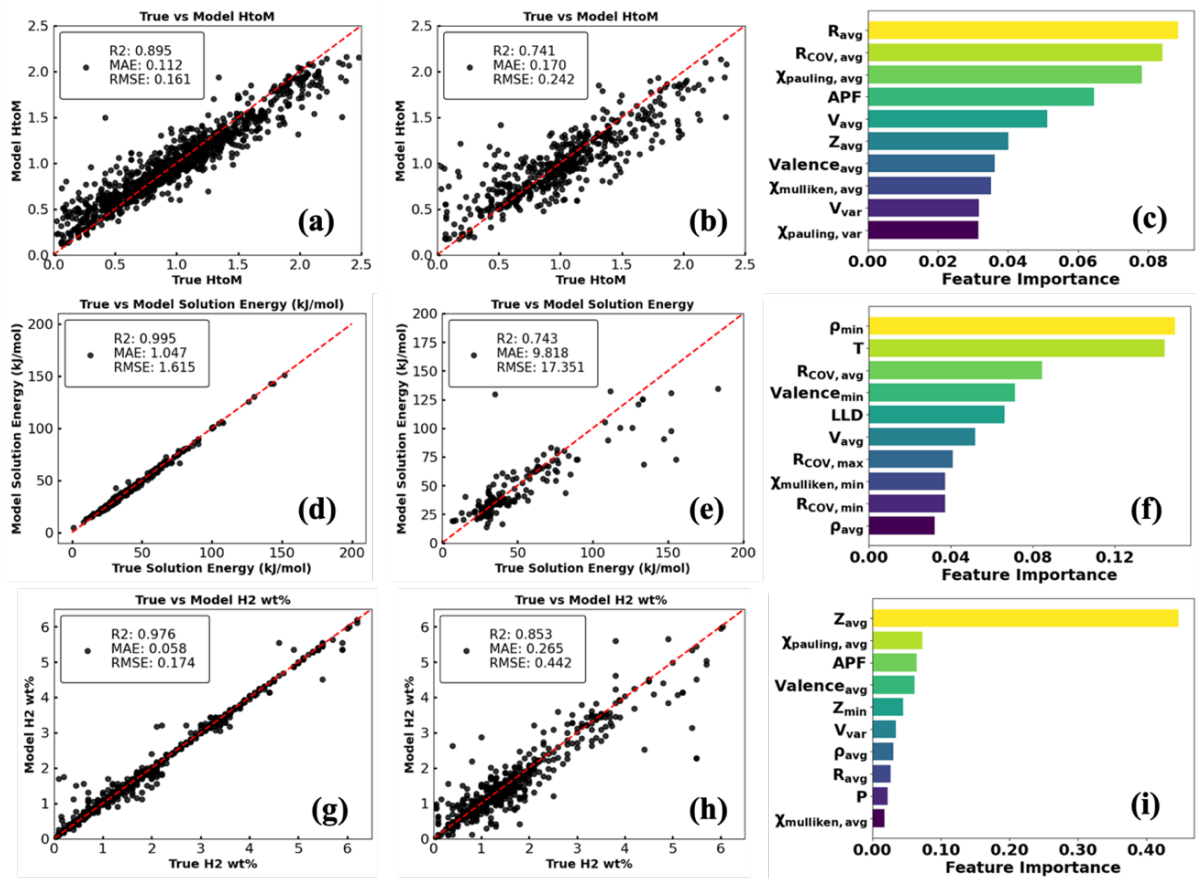

**Figure 3.** Parity plots and feature importance results for the 3 models: (I)-H/M, (II)-Solution energy, (III)-$H_2$ wt%, where (a) training (b) test data parity plots, (c) top 10 important features for the best performing models in each of those three cases.

Additionally, the $H_2$ wt% model fits better (>85% $R^2$ on test data) than the H/M model (~74% $R^2$ on test data), indicating the strong effect of $Z_{avg}$ and the model is able to capture the complexity and variance of $H_2$ wt% with the selected feature set in the dataset. As mentioned before, among the 1775 data points only 558 compositions of all hydride types (AB, AB2, AB5, MIC, and SS) have solution energy, and therefore for the solution energy ($\Delta H$) predictive model only these 558 data points are used and **Figure 3d-f** shows the training, test data parity plots, and feature importance of the $\Delta H$ model. It is found that other than $R_{COV}$; $\rho_{min}$, temperature, and LLD are among the top 5 feature sets. Our ML model for solution energy attains an $R^2$ score of 74 % and a 5-fold cross-validation $R^2$ of 75% over the unseen dataset based on the XGB model which shows the best fitting and accuracy. As $\Delta H$ shows strong linear dependencies with equilibrium pressure and temperature (T), given by Van't Hoff's relation [58–60] shown in equation (7), therefore the temperature-dependency is relevant.

$$\ln\left(\frac{P_{eq}^0}{P}\right) = -(\Delta H)\left(\frac{1}{RT}\right) + \frac{\Delta S}{R} \qquad (7)$$

This comes from replacing the expression of $\Delta G$ in $-RT \ln K = \Delta G$ by $\Delta G = \Delta H - T\Delta S$, where $K = \left(\frac{P_{eq}^0}{P}\right)$, is the equilibrium constant from the Pressure-Composition-Isotherm (PCI) of metal alloys for hydrogen storage, $\Delta G$, $\Delta H$, and $\Delta S$ are Gibbs free energy upon hydrogenation, change in standard enthalpy or heat of formation during hydrogenation and change in net entropy respectively, $P_{eq}^0$ is the equilibrium pressure at 25°C, R is the universal gas constant.

### 3.2 Feature importance on H/M and solution energy predictions

Our predictions from the ML model are closely related to existing literature, which shows that hydrogen adsorption and desorption increase significantly with the addition of Ti in MPEAs [61,62]. It has also been reported that Ti and Zr exhibit some of the highest hydrogen uptake capacities among metals [63,64]. Our ML results, including models like RF for H/M, XGB for solution energy, and GBR for $H_2$ wt%, indicate that covalent radius (RCOV), average electronegativity ($\chi_{pauling}$), and average atomic radius ($R_{avg}$) are the most critical features, as shown in **Figure 3c-i**. These findings are consistent with literature, where covalent radius, LLD, and electronegativity are known to be key determinants in hydrogen adsorption capacity in MPEAs due to atomic and electronic structure considerations [65,66].

Interestingly, we found that LLD appears as one of the top 5 important features in the solution energy predictive model (**Figure 3f**), but adding LLD did not significantly affect the H/M and $H_2$ wt% predictions. This may suggest that while lattice distortion (LLD) plays an important role in the solution energy model, it does not directly influence the hydrogen uptake metrics in our ML predictions, as expected based on previous findings that high lattice strains from distortion are favorable for hydrogen adsorption in high-entropy alloys and alloys with low VEC [18,67]. On the other hand, when looking at hydrogen weight percentage prediction in Figure 3i, the top 5 features in the GBR model—atomic weight (Z), atomic-packing factor (APF), $\chi_{Pauling}$, and average valence—contribute significantly (60-70%) to the prediction. Interestingly, atomic weight (Z) was not among the top 5 features in the H/M predictive model. Both covalent radius ($R_{COV}$) and electronegativity ($\chi_{Pauling}$) consistently emerged as significant contributors across all tree-based models, highlighting their central importance in H/M prediction. We also noted that Vanadium and Titanium are two crucial elements that may improve H/M on alloying [61,65]. In **Supplementary Figure S1**, we have shown the actual vs predicted data points of both H/M and Solution Energy models with the number of data entries in continuous plots using 4 different ML models: GBR, RF, XGB, and Ridge regressions. In **Supplementary Figure S2**, we have shown the effect of binary equiatomic refractory alloy in their H-storage capacity which shows that of Nb-V, Nb-Ti, and Nb-Zr show the best H-storage capacity, attaining a H/M of 1.38, 1.24, and 1.27 respectively. These relatively higher H/M values underscore the potential of these alloys for effective H-storage applications.

**3.3 Effect of Different Elemental Compositions on ML developed Predictive Models**

The BCC materials have been studied extensively for H-storage application as it is often identified to store H/M = 1 (monohydride) in a BCC or body-centered tetragonal (BCT) and H/M = 2 (dihydride) in face-centered cubic (FCC) phase [69–72]. However, hydrogen addition either destabilizes the alloy's hydride phase by either decreasing or increasing the stability, which is not conducive to hydrogen desorption [73–76]. Thus, in this section, we analyzed the H/M behavior of single-phase binary bcc refractory alloys using our GBR-predicted ML model to understand the role of refractory metal elements.

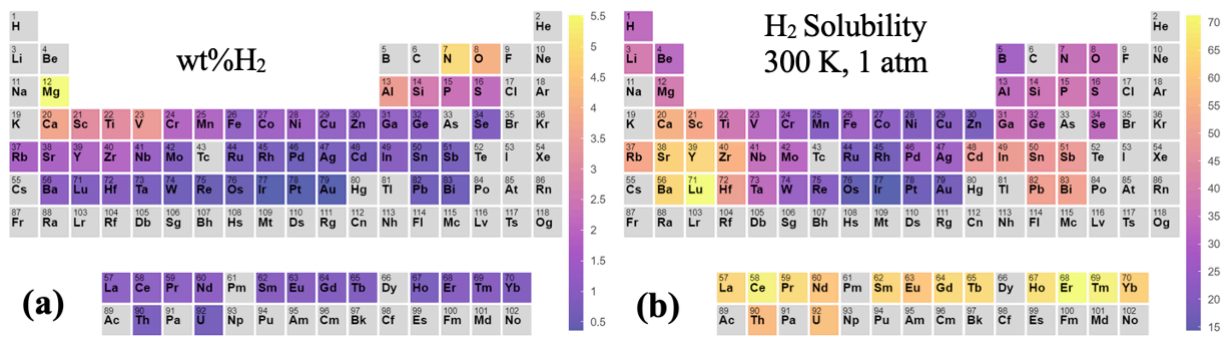

**Figure 4.** Periodic tables illustrating elemental effects on (a) $H_2$ wt% and (b) solution energy, derived from ML model predictions, which provides a reference for identifying alloying elements that enhance H solubility and storage.

**Figure 4a-b** shows the effect of binary elements on their ability to store hydrogen in terms of hydrogen weight percent and solution energy results based on our ML predictive model. Among the transition metal elements, V, Nb, Ti, and Zr show high H-storage ability with binary vanadium showing a 3.0-3.5 wt% H-storage ability, whereas the presence of other transition metals like Mo, Ta, and W shows hydrogen storage of only about 0.5-1.0 wt%. This is consistent with literature showing that V, Ti, Nb, and Zr have high affinity towards hydrogen and form stable solid solutions of hydrides but Mo, Cr exhibit low hydrogen affinity, leading to lower hydrogen storage [26,77]. Additionally, experimental results show that magnesium-based alloys exhibit 5.0-5.5 wt% hydrogen storage. However, magnesium can store hydrogen of about 6.0-6.4 wt % at ambient pressure and >350°C.

**3.4 Ternary Ti-Nb-X (X=Mo, Cr, Hf, Ta, V, Zr) MPEAs**

Based on the results shown in **Figure 4**, in this section, we looked into the H-storage ability of transition and refractory alloy compositions having different compositions. We chose different ternary compositions of Ti-Nb alloys with Mo, Cr, V, Zr, Hf, and Ta to denote the effect of different alloying compositions on their H-storage ability, as shown in **Figure 5**.

From the H/M ternary plots, it is clearly seen that the Ti-rich and Ti-V-rich regions of all the compositions show a hydrogen adsorption capacity of more than 2.4-3.5 wt%. In most of the compositions, higher hydrogen weight percent regions also exhibit higher H/M ratio, however, for TiNbHf and TiNbZr 80-100% Ti-rich regions predict the highest $H_2$ storage of 2.5-3.0 wt% whereas the Hf- and Zr-rich regions show the highest H/M ratio based on our ML model's prediction. Hf, Zr, and Ta- all three have a very high molecular weight (>90 g/mol), therefore, even though the ML model predicts their high affinity towards hydrogen, the hydrogen storage in Hf-, Zr-, and Ta-rich regions remains only 1-2 wt%. It agrees well with our earlier results showing average molecular weight is the most dominant feature (>40% effect) in predicting $H_2$ weight percent (**Figure 3i**). This result also reflects the fact that the $H_2$ weight percent predictive model gives a more relevant understanding of overall H-storage capacity than the estimation of H/M. Additionally, these ternary plots allow for a comprehensive and systematic exploration of the compositional space. By visualizing the effects of varying concentrations of different transition elements, the predictions of H-storage capacity and stability can be further analyzed. This approach minimizes the need for exhaustive experimental testing by focusing efforts on the most promising compositions.

The model predictions in **Figure 5** show that Nb-Ti-based MPEAs are expected to have higher H/M and, therefore, a better fit for hydrogen storage applications. This finding corroborates with earlier reports [78] where early transition metals show improved stability for hydride formation while it rapidly decreases moving across the period for mid-to-late transition metals. Notably, Ti-based alloys also present several benefits including fast H-storage, long cycle life, better release rate, and low price, thus, it is considered one of the best materials choices for hydrogen storage. The choice of the third element, i.e., molybdenum, was based on prior knowledge that Mo-substitution can extend the BCC stability range. The superior ability of Nb-Ti alloys doped with Mo in stabilizing the BCC structure comes from reduced unit-cell volume or lattice parameter in the BCC phase (due to smaller atomic radii of Mo-atom compared to Ti-atom) that inhibits the transformation of Ti into HCP phase. The coarse packing and high density of interstitial sites (tetrahedral and octahedral) in the BCC structure also make these alloys attractive for superior hydrogen absorption.

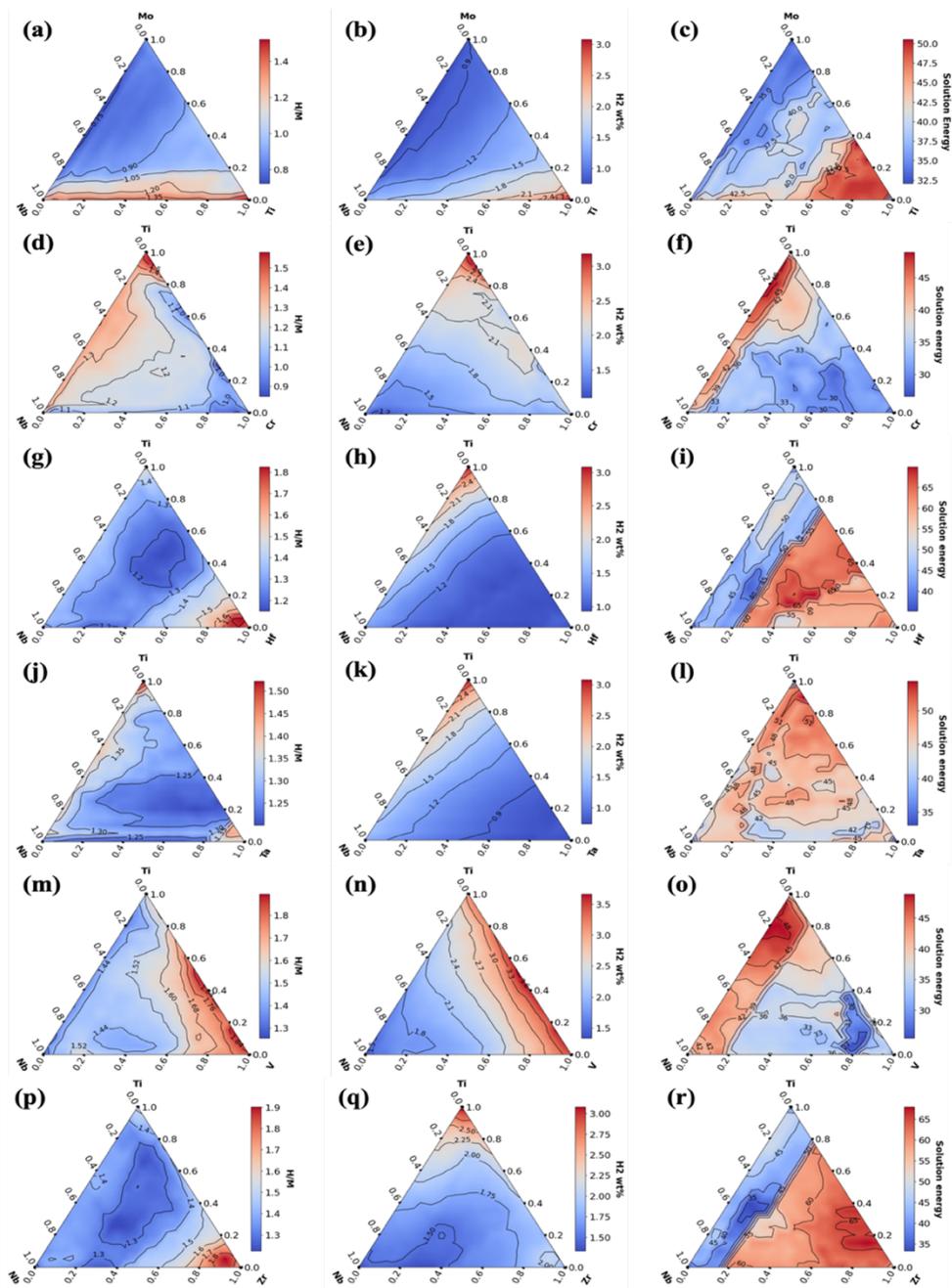

**Figure 5.** Ternary plots of different transitional metal compositions showing the distribution of H/M, hydrogen weight percent and solution energy. (a-c) TiNbMo, (d-f) TiNbCr, (g-i) TiNbHf, (j-l) TiNbTa, (m-o) TiNbV, (p-r) TiNbZr. (a,d,g,j,m,p) are for the H/M, (b,e,h,k,n,q) are for hydrogen weight percent and (c,f,I,l,o, and r) are for the solution energy based on our ML models predictions.

The ternary plots indicate that the combination of Ti, Nb, and Mo yields a favorable H/M ratio with a stable Nb-Ti rich solution energy region. The solution energy plots showed that the Ti-Nb-Mo combination has a relatively low solution energy, indicating stable hydride formation. Stability is essential for the reversibility of hydrogen absorption and desorption, making this composition a reliable candidate for hydrogen storage. Additionally, the combination of Mo, Nb, and Ti offers a good balance between thermodynamics (solution energy) and kinetics (hydrogen absorption) which is critical for H storage.

**3.5 Phase stability and electronic-structure analysis of Ti-Nb-Mo Alloys**:

To further understand the phase stability and hydrogen storage behavior, in **Fig. 6a**, we show the ternary diagram of formation enthalpy for the bcc Ti-Nb-Mo alloy. The phase stability diagram reveals two distinct thermodynamic regions: (1) the high-entropy region, located at the center of the ternary diagram, which shows higher stability, and (2) the vortex regions or edges, corresponding to binary or near-unary regions, which exhibit weaker stability.

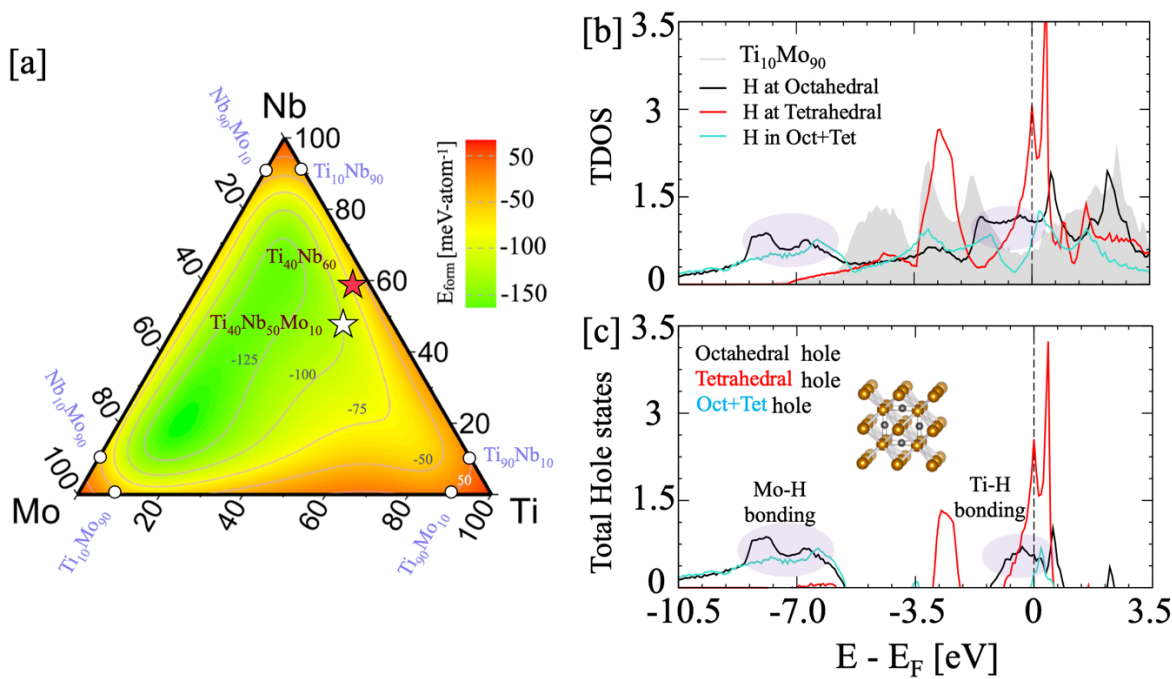

**Figure 6**. (a) Thermodynamic stability (formation enthalpy) analysis of ternary BCC Ti-Nb-Mo MPEAs. (b) The electronic-structure analysis of Mo-rich $Mo_{90}Ti_{10}$ shows strong Mo-*d* and H-*s* bonding at very low energy below Fermi-level, and (c) total hole states where H-*s* goes are near very deep energy states, i.e., -7.0 eV, not very favorable for hydrogen storage or absorption/desorption cycle. See related discussion on relaxation effect on DOS of Mo-Ti alloy in **Appendix A1 and A2** and **Table A1**. While the experimental validation of detrimental role of Mo on hydrogen absorption is discussion in **Appendix A3**.

In the near-equiatomic region, the formation enthalpy of the bcc phase sharply increases. Interestingly, the Mo-Nb-rich region is thermodynamically more stable, while the Ti-rich region shows moderate to low phase stability. From a hydrogen absorption perspective, the high stability of hydride formation in the Mo-Nb region makes the desorption process slower and requires more energy to break the metal-hydrogen bonds. This is consistent with the observation that very high VEC metal atoms are less favorable, as the contraction of metal-d bands across the period reduces the extent and overlap of the d-states [78]. The phase stability data from DFT further confirms the computational insights gained from the ML models, offering a detailed understanding of the thermodynamic and structural factors influencing hydrogen storage performance in these alloys.

The ternary diagram in **Fig. 6a** shows the DFT calculated phase stability (formation enthalpy) diagram for bcc Ti-Nb-Mo alloy. The phase stability diagram can be divided into thermodynamic regions: (1) high-entropy region, i.e., the central part of the ternary diagram, with higher stability, and (2) vortex regions or edges, i.e., binary or near unary regions, with weaker stability. The formation enthalpy of the bcc phase increases sharply in the near equi-atomic region. Interestingly, Mo-Nb rich region is thermodynamically more stable while the Ti-rich region shows moderate to low phase stability, which is interestingly from a hydrogen absorption point of view because the very high high-stability of hydride formation makes the desorption process slow as it will require higher energy to break metal-hydrogen bonds. The reason for not choosing very high VEC metal atoms lies in the contraction of metal-*d* bands going across the period due to increased nuclear charge that reduces the extent and overlap of the *d*-states [78].

In **Fig. 6b**, we show the electronic structure (electronic density-of-states) of binary bcc $Ti_{10}Mo_{90}$ and its hydride where hydrogen is added at octahedral (O; 3 sites), tetrahedral (T; 6 sites), and mixed octahedral-tetrahedral (O+T; 3+6 sites) sites as shown in the inset. The grey-shaded region is the total DOS of $Mo_{90}Ti_{10}$ without hydrogen addition while black, red, and cyan represent TDOS with hydrogen at three different interstitials. The H-*s* states for octahedral and tetrahedral hydrogen occupation are mainly concentrated in low-energy {-6 to -10 eV} and mid-energy {-3 eV} to high at/around Fermi-level, respectively, where the Fermi-level is defined at zero on the x-axis. The **Fig. 6c** shows the hole-states added due to hydrogen addition to alloy, which was calculated by subtracting TDOS of pristine $Ti_{10}Mo_{90}$ with TDOS of hydrogenated alloys or by employing the analytical relationship, i.e., $D_h(\varepsilon) = \frac{dN}{d\varepsilon} = \frac{V}{2\pi^2} \times \left[\frac{2m_h}{(h/2\pi)^2}\right]^{3/2} \times [\varepsilon + E_F]^{1/2}$, where $N(k) = \frac{k^3 V}{3\pi^2}$, and $k^2 = 2 \times m_h(\varepsilon + E_F)/(h/2\pi)^2$ where N is the number of electronic states, V is the unit-cell volme, $m_h$ is hole mass, and $E_F$ is the Fermi-energy.

The TDOS of $Ti_{10}Mo_{90}$+H was scaled by the difference of percent lattice change for one-to-one comparison. We observed in **Fig. 6b,c** that the low energy bands in $Ti_{10}Mo_{90}$+H show pronounced hydrogen character. Additionally, a metal character arising from the formation of bonding and antibonding states between the metal $e_g$ and H-s states was also observed. The observation of higher density of H-*s* states from tetrahedral or mixed (O+T) hydrogen near/at the Fermi-level (see **Fig. 6b**) suggests that these states will require lower energy to break metal-hydrogen bonding for hydrogen desorption. Despite the higher density of hydrogen hole states in the tetrahedral case over the octahedral case, a peak in the TDOS at the Fermi level adds a key drawback, i.e., energy instability, which has been reported in the past in other refractory metal alloys [79]. Interestingly, the hydrogen addition at the mixed O+T sites in {Ti, Nb} shows no such instability, thereby, suggestive of a higher concentration of hole-states, which could be advantageous for improved hydrogen absorption.

In **Fig. 7a-i**, we present a systematic analysis of electronic DOS of all alloying compositions predicted from GB-based ML model as well with constituent elements {Nb, Ti, Mo} to understand the electronic-structure origin of improved hydrogen storage behavior. If we carefully look at the electronic DOS of {Nb, Ti, Mo} metals, in **Fig. 7a-c**, we can clearly see that low VEC refractory metals such as Ti show higher spread in electronic states of metal-*d* bands while high VEC shows stronger localization of *d*-states, e.g., Nb and Mo. Notably, titanium is known to be a better hydride-forming element compared to other transition metals, and remarkably, the electronic density of states of the Ti-H system in **Fig. 7b** shows that the population of H-s states is much closer to Fermi-level (-3 to -7 eV) compared to Mo-H (-5 to -9 eV) and Nb-H (-4 to -8 eV).

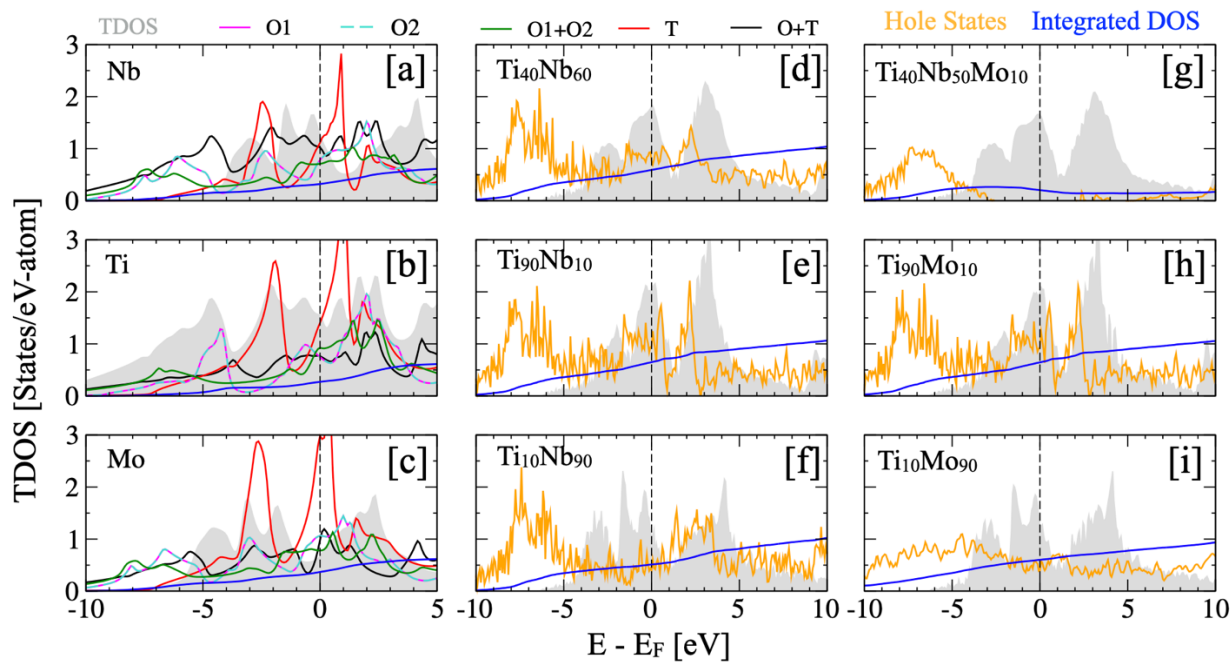

**Figure 7.** (a-c) Total electronic density of states (TDOS) of simple metal {Nb, Ti, Mo}-hydrides in the body-centered cubic structure. (d-i) TDOS and hole-states (considering octahedral hydrogen) in six ML (GB) predicted Ti-Nb-Mo MEAs. The solid blue line shows the trendline for the integrated density of states.

The ML models in **Fig. 2 and 3** show a somewhat positive correlation of LLD with hydrogen-to-metal ratio (H/M) and hydrogen solubility in hydrogen storage materials. To extract LLD, in this section, we employed the first-principles DFT method to fully optimize (atomic and lattice) the disorder supercell of chosen compositions and utilized a revised definition of atomic distortion as $\Delta u_i^{rlx-ideal} \sim \sqrt{|u_{rlx}^i(x,y,z) - u_{ideal}^i(x_0,y_0,z_0)|^2}$ or $\sqrt{(x-x_0)^2 + (y-y_0)^2 + (z-z_0)^2}$, which could be used for both crystalline, partially-order or disorder solids [83,84]. Here, $u_{rlx}^i(x,y,z)$ and $u_{ideal}^i(x_0,y_0,z_0)$ are positions of $i^{th}$ atom at relaxed and special lattice symmetry points. The atomic displacement of each atom in the crystalline lattice was then used to estimate the vector ($L_{2,1}$) norm of atomic displacements, i.e., LLD. The proposed approach was utilized to estimate the LLD, which is tabulated in **Table 3**. The comparison between ideal and relaxed supercell shows that LLD has a strong effect on formation enthalpy ($E_{form}$), which is further highlighted in $\Delta E_{form}^{ideal-rlxd}$. Our calculations show that a strong LLD contribution in Ti-Nb-Mo significantly influences the phase stability and affects the material's overall hydrogen storage capacity. We found only two predicted compositions, i.e., $Ti_{40}Nb_{60}$, and $Nb_{40}Ti_{50}Mo_{10}$, fall in optimal formation enthalpy region, which also show higher LLD. Both phase stability and LLD suggest improved hydrogen storage, however, looking back into electronic-structure $Ti_{40}Nb_{60}$ (**Fig. 7d**) and $Ti_{40}Nb_{50}Mo_{10}$ (**Fig. 7g**), we found an absence of hole states in $Ti_{40}Nb_{50}Mo_{10}$ compared to $Ti_{40}Nb_{60}$, which is indicative of lower hydrogen absorption.

**Table 3:** DFT calculated local-lattice distortion, formation-enthalpy, formation enthalpy difference, and supercell volume for selected set of BCC Ti-Nb-Mo alloys (also see **Table A1** and related discussion in **Appendix A2**).

| System | LLD [RLX-Ideal] | $\Delta E_{form}$ [meV/atom] | | $E_{form}$ [Ideal - RLX] | V [Å³/atom] |
|---|---|---|---|---|---|
| | | Ideal | Relaxed | | |
| $Ti_{90}Mo_{10}$ | 0.026 | -127.4 | -191.1 | 63.8 | 16.6 |
| $Ti_{10}Mo_{90}$ | 0.171 | -60.8 | -62.4 | 1.7 | 15.7 |
| $Nb_{90}Mo_{10}$ | 0.193 | +10.7 | +10.2 | 0.6 | 17.9 |
| $Nb_{10}Mo_{90}$ | 0.038 | +19.9 | +16.7 | 3.1 | 15.9 |
| $Ti_{90}Nb_{10}$ | 0.216 | -101.9 | -132.9 | 31.0 | 17.1 |
| $Ti_{10}Nb_{90}$ | 0.034 | +101.8 | +99.9 | 1.8 | 18.03 |
| $Ti_{40}Nb_{60}$ | 0.121 | -11.9 | -31.4 | 19.5 | 17.49 |
| $Ti_{40}Nb_{50}Mo_{10}$ | 0.124 | +1.5 | -30.6 | 32.0 | 17.48 |

Notably, our analysis suggests that the addition of LLD as a feature gives us a better understanding and ability to predict the hydrogen storage abilities of these chemically complex materials. The stability of MPEAs is also dominated by the atomic and electronic features of their constituent elements in the solid solution including electronegativity, electron affinity, atomic weight, volume, radius, and covalent bond at different temperature and pressure conditions. We have shown in **Fig. 4** the importance of including these features in predicting the hydrogen storage performance of metal hydrides [77]. Finally, we also want to emphasize the significance of LLD in the refractory alloys [85] as the atomic size mismatch, which is systematically used as LLD qualifier, however, it is not accurate enough to predict LLD. In general, a higher LLD in alloys means additional lattice strain, which is seen as a useful feature for quantifying hydrogen adsorption as well as hydride formation [31,67].

**3.6 Comparative analysis with molecular simulations**

**Figure 8** shows molecular simulation results for different binary and ternary systems of TiNbMo-based alloys to qualitatively assess hydrogen diffusion at a temperature of 1000 K. Simulations were carried out under two conditions for each composition, one with ideal unrelaxed crystal structure and the other where the atomic arrangement of the structures are relaxed to find its minimum energy configuration.

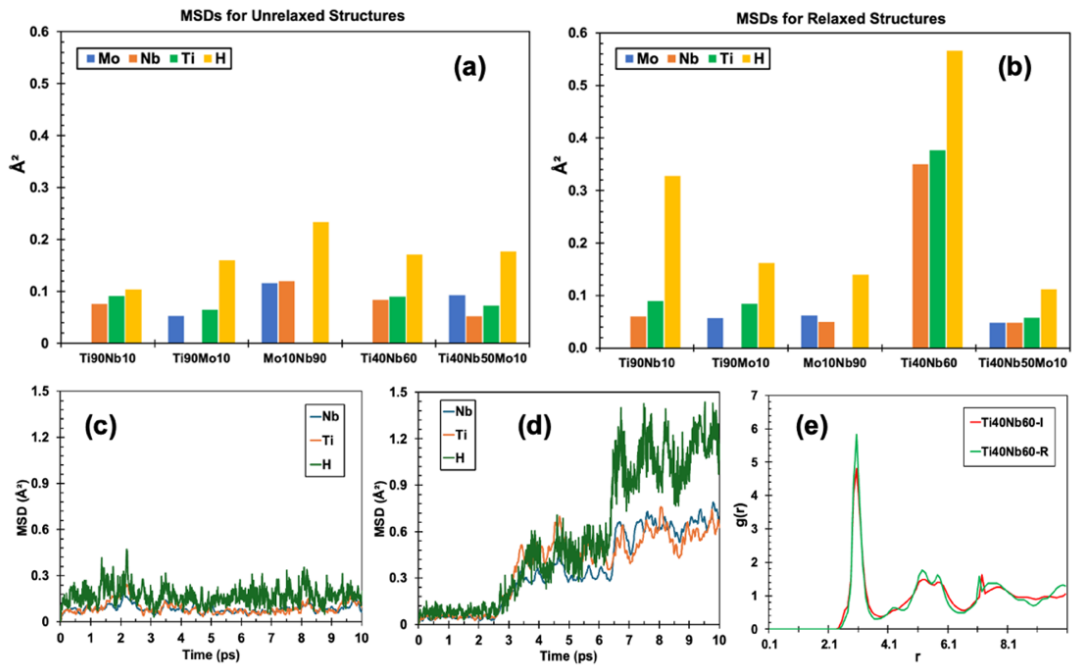

**Figure 8. (a-b)** Mean square displacements (MSD) of hydrogen and different Ti-Nb-Mo systems in the presence of hydrogen. **(c-d)** MSDs of elemental compositions in hydrogen environment for **(c)** unrelaxed, **(d)** relaxed structures of $Ti_{40}Nb_{60}$, and **(e)** radial distribution functions of both cases ($Ti_{40}Nb_{60}$-I refers to the unrelaxed and $Ti_{40}Nb_{60}$-R refers to the relaxed structure respectively).

**Figure 8 (a-b)** shows the average mean squared displacements (MSD) of elemental compositions and molecular hydrogen for both unrelaxed and relaxed structures of $Ti_{90}Nb_{10}$, $Ti_{90}Mo_{10}$, $Nb_{90}Mo_{10}$, $Ti_{40}Nb_{60}$ and $Ti_{40}Nb_{50}Mo_{10}$ systems. This clearly shows that 'Ti' shows a positive effect in hydrogen diffusion for relaxed binary solutions of different weight percent of Ti-Nb systems compared to the unrelaxed ones suggesting that LLD induced during relaxation plays quite an important role in hydrogen diffusion in these structures whereas all the other 3 combinations involving 'Mo' ($Ti_{90}Mo_{10}$, $Nb_{90}Mo_{10}$ and $Ti_{40}Nb_{50}Mo_{10}$) shows a 20-30% reduction in hydrogen diffusivity i.e., the involvement of 'Mo' downplays the LLD effect. We have also provided the discussion on role of relaxation on electronic-strucure, which corroborates with findings of MD where relaxation changes diffusion kinetics.

**Figure 8 (c-e)** depicts the time evolution of average elemental MSDs and radial distribution functions for the $Ti_{40}Nb_{60}$ structure. Additionally, when 10% of Nb in $Ti_{40}Nb_{60}$ is replaced with 'Mo' to form $Ti_{40}Nb_{50}Mo_{10}$, average hydrogen MSD for the relaxed structure (as in **Figure 8 (b)** and **Supplementary Figure S3(a-b)**) shows a six-fold increase than for $Ti_{40}Nb_{60}$, although, the MSDs for the unrelaxed structures are similar for these 2 compositions. These findings are consistent with our results obtained from the ML model (**Figure 3**), suggesting that increasing "Mo" shows a negative impact on H/M (see Appendix **Fig. A2** for experimental valdiation). **Supplementary Figure S4** also shows that for both the $Mo_{10}Nb_{90}$ and $Mo_{90}Nb_{10}$, the relaxed structures have lower average hydrogen MSDs which is contrary to the $Ti_{40}Nb_{60}$ (**Figure 8 (c-d)**). This suggests that the lattice distortions induced by the atomic positional reorientation restrict the interstitial diffusion of hydrogen inside the lattice structure, despite maintaining structural integrity, as observed in the radial distribution functions (RDF) in **Supplementary Figure S4 (c)** and **(f)**. **Supporting figures S5** and **S6** show the corresponding MSD and RDF of other structures ($Ti_{10}Nb_{90}$, $Ti_{90}Nb_{10}$, $Ti_{10}Mo_{90}$, and $Ti_{90}Mo_{10}$) analyzed in the study and **Table S1** shows all the average MSD values of each of the elemental compositions averaged over 10 picoseconds (ps) for different Ti-Nb-Mo systems in the presence of hydrogen.

**3.7 CoNi-X (X=Al, Mg, V) systems**

In this section, we discuss different FCC ternary compositions of Co-Ni-X alloys (**Figure 9**) with X=Al, Mg, and V to denote the effect of different alloying compositions on their hydrogen storage ability.

From the H/M ternary plots, it is found that the V-rich (V > 80%) zone in CoNiV (**Fig. 9g-i**) and Co-Ni-rich regions of Co-Ni-Al (**Fig. 9a-c**) compositions shows a hydrogen adsorption capacity of more than 1.8-3.0 wt%. However, for the CoNi-Mg (**Fig. 9d-f**), for Mg content of >70% in the solid solution, our model predicts an H/M of 1.4-1.8 and correspondingly an $H_2$ weight percent of 4%. This is because, with increasing Mg content, the alloy becomes increasingly more stable at the HCP phase having larger interstitial gaps for hydrogens to occupy. Additionally, by increasing the Mg and Al content beyond 60% in CoNiMg and CoNiAl systems, the solution energy formation becomes more favorable, and the hydrogen weight percent is 1.5-4%. However, excessive Al content can also lead to grain coarsening and alter the solidification path, potentially affecting hydrogen storage properties and the presence of high-Mg contributes to high hydrogen weight percentages [79]. This also suggests that while Al and Mg play a crucial role in the thermodynamical stability of the alloy after hydrogenation, the presence of Co and Ni contributes to the formation of intermetallic compounds (like $Mg_2Ni$) having better hydrogen adsorption and desorption kinetics. The results suggest a trade-off between hydrogen storage capacity and thermodynamic stability while V-rich compositions show high H/M ratios, Mg-rich compositions offer more favorable solution energy formation [80,81]. Therefore, based on these results, 3 promising compositions are identified to perform an atomistic study on the hydrogen diffusion in bulk alloy phases starting with the initial FCC phases of $Ni_{70}Co_{20}Al_{10}$, $Ni_{10}Co_{10}V_{80}$, and HCP phase of $Ni_{15}Co_{10}Mg_{75}$.

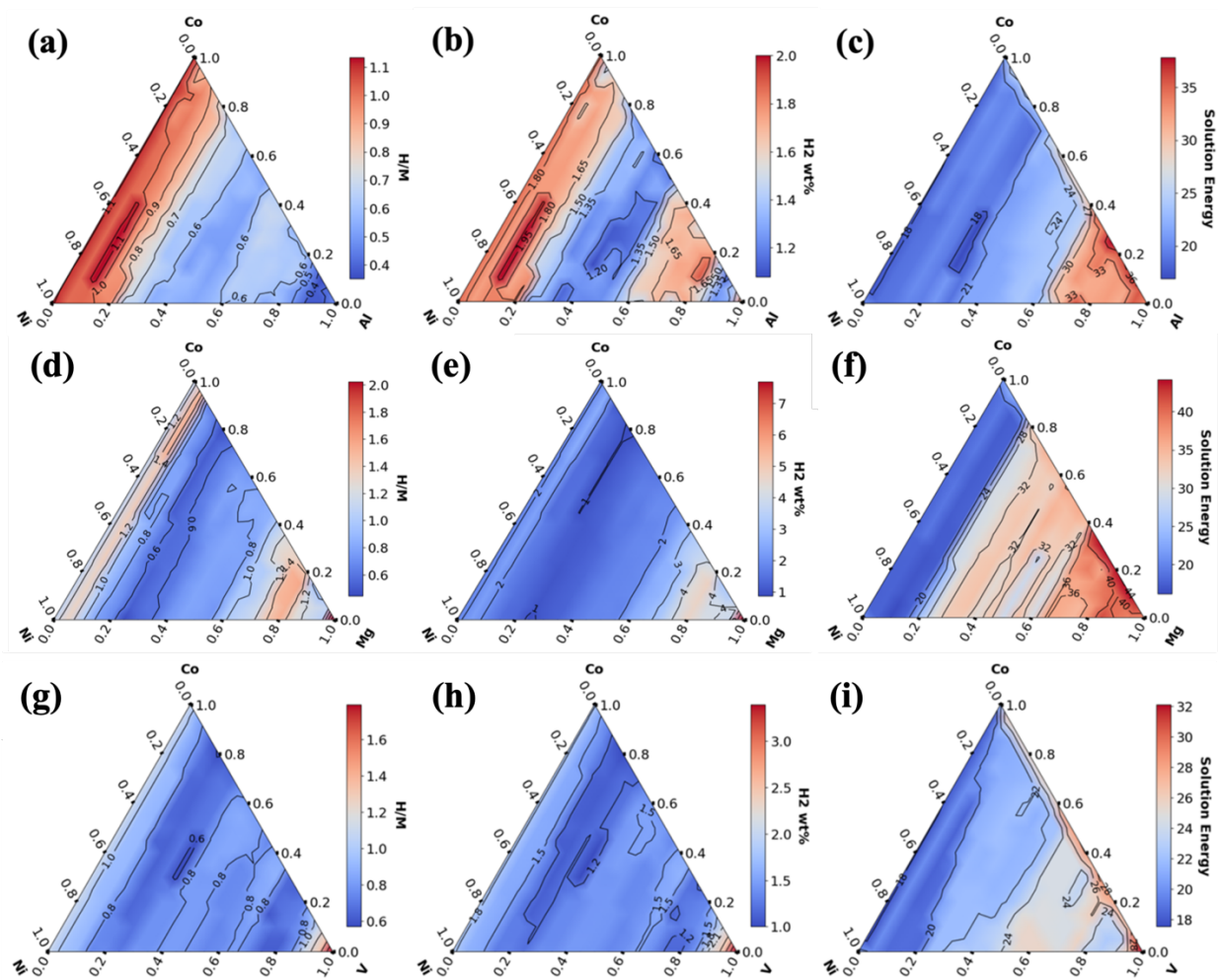

**Figure 9.** Ternary plots of different transitional metal compositions show the distribution of H/M, $H_2$ wt %, and solution energy. (a-c) CoNi-Al, (d-f) CoNi-Mg, (g-i) CoNi-V. (a, d, g) are for the H/M, (b, e, h) are for hydrogen weight percent, and (c, f, i) are for the solution energy based on our ML models predictions.

Based on the ML predictions of different CoNi-X (X=Al, Mg, V) ternary systems, here we present an atomistic insight of hydrogen diffusion characteristics in 3 chosen alloys. **Figure 10** shows the average logarithmic MSDs over 1 ns of hydrogen diffusion simulation at 2 different temperatures on the 3 selected

CoNi-based ternary compositions. The actual average MSDs are shown in **Table S2** in **Supplementary information**. **Figure S5** of supplementary information shows the average MSDs of elemental compositions with hydrogen for the three different compositions analyzed in this study and the corresponding radial distribution functions show the structural information after hydrogen diffusion within the metal matrix. We observe a striking 2 orders of magnitude higher average hydrogen mean squared displacements in Mg-based alloy than the V-rich or N-rich Al alloy establishing our previous observation from the ML study (**Figure 9(d-f)**) that Mg has a strong positive effect on the hydrogen storage capacity and helps in thermodynamic stability of ternary CoNi-Mg alloy. Finally, the thermodynamic stability of CoN-X systems enables multi-principal element alloying without phase separation, ensuring a robust host lattice for hydrogen absorption. Local lattice distortion or average mean square displacements in Co-Ni alloys is a key factor that allows controlled distortions, which enhance hydrogen binding without excessive trapping.

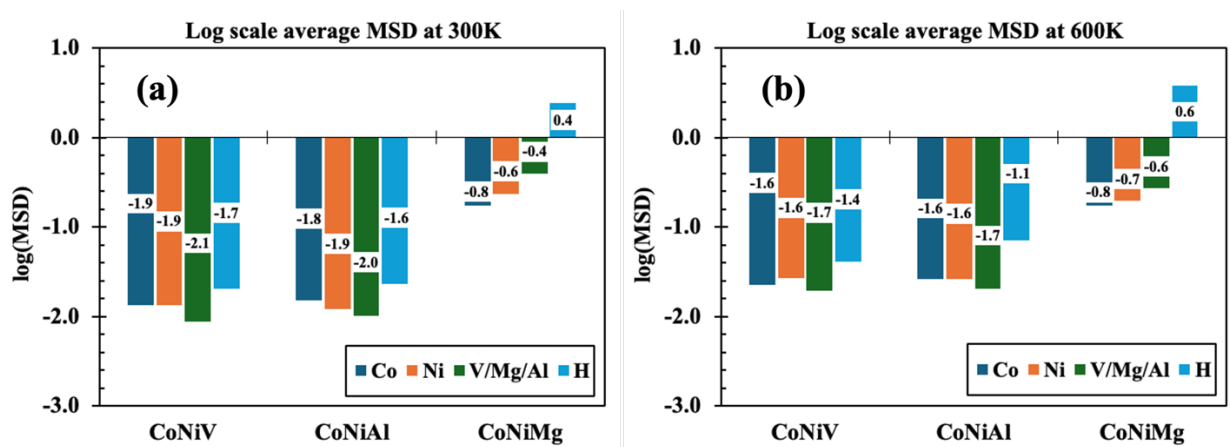

**Figure 10. (a-b)** Average mean square displacements (MSD) of elemental compositions in the presence of hydrogen for different CoNi-X ternary systems (in log-scale). The table shows the MSDs of hydrogen in these 3 different systems. The HCP structure of Mg-rich CoNi-Mg exhibit almost 2 orders of magnitude higher hydrogen MSD at both the different temperatures than the V- and Al- ternary systems.

## 4. Summary

We performed comprehensive feature engineering to identify key properties governing the hydrogen-to-metal (H/M) ratio and solution energy in chemically complex solid-solution alloys. Local lattice distortion (LLD) is a critical factor in refractory-based alloys, influencing both solute miscibility (e.g., H) and mechanical properties. To capture its effect, we incorporated atomic, electronic, and lattice features in our machine learning (ML) models for predicting H/M and hydrogen solubility. Our results highlight that atomic radius, covalent radius, and electronegativity, along with lattice features such as atomic packing fraction (APF) and LLD, dominate predictive outcomes across ensemble-based ML models. Notably, LLD significantly impacts solution energy predictions, with our best-performing model, XGBoost (XGB), achieving an R² of 74% and a mean absolute percentage error (MAPE) below 10% on test data, alongside a 75% average R² in cross-validation. Additionally, atomic weight, packing fraction, electronegativity, and valence collectively contribute over 60% to hydrogen weight percent predictions in Gradient Boosting Regression (GBR) and XGB models. Temperature and average density emerge as key parameters for solution energy predictions, aligning with Van't Hoff equation for thermodynamic equilibrium.

To validate our models, we tested multiple alloy compositions, revealing that Ti, Nb, and V enhance H-storage capacity across diverse compositions. However, adding just 10-20% Mo reduces H/M and H wt% by 40-50%. Specifically, predictions for the Ti-Nb-Mo system indicate that increasing Ti and Nb content can elevate H/M to ~1.4 and hydrogen storage to ~3.0 wt%, whereas Mo negatively impacts storage performance. Molecular simulations confirm these trends, showing that Ti and Nb promote H diffusion, while Mo hinders it. Furthermore, DFT reveals that hydrogen preferentially hybridizes with Mo *d*-orbitals at low energies below the Fermi energy, making Mo unfavorable for hydrogen storage. Phase stability and lattice distortions in bcc Ti-Nb-Mo alloys significantly influence their H-absorption behavior. Despite moderate to low phase stability in Ti-rich regions, favorable Ti-H bonding dynamics enhance H uptake. Incorporating LLD in ML models improves prediction accuracy, underscoring the interplay between electronic structure and atomic displacements in these complex alloys. Finally, the ML-derived periodic table on (a) hydrogen wt% and (b) solution energy will further aid material selection by identifying alloying elements that enhance hydrogen solubility and storage. These findings provide a framework for selecting optimal H-storage materials and guiding future experimental efforts.

## 5. Acknowledgments

This work was supported by the Laboratory Directed Research and Development (LDRD) program at Ames National Laboratory. Kevin Ji is grateful for the research opportunity at Ames National Laboratory supported by the U.S. Department of Energy (DOE), Office of Science, Science Undergraduate Laboratory Internships (SULI) program. The Ames National Laboratory is supported by the U.S. Department of Energy (DOE), Office of Science, Basic Energy Sciences, Materials Science and Engineering Division. The Ames National Laboratory is operated for the U.S. DOE by Iowa State University under contract DE-AC02-07CH11358. TB and GB acknowledge the support from the National Science Foundation awards CMMI-1944040 and CMMI-2436601, respectively. PS would like to thank Drs. Matthew D. Wittman, Mark D. Allendorf, and Vitalie Stavila of Sandia National Laboratory for useful discussion and providing updated HyMark database.

**6. APPENDIX**:

**Appendix 1.1 Local lattice distortion and structural stability: Table A1** provides a detailed analysis of LLD, phase stability, and volume change in body-centered cubic (BCC) multi-principal element alloys (MPEAs), particularly focusing on the effects of hydrogen incorporation and oxygen impurities. The formation energy ($E_{form}$), percentage change upon relaxation, and equilibrium volume per unit cell and per atom ($V_0$) are examined across different alloy compositions. For pure Mo, the formation energy is relatively low at 25.9 meV/atom, indicating stability, while the atomic volume per atom is 15.78 Å$^3$. Upon hydrogen incorporation (Mo+H), the formation energy increases drastically to 624.5 meV/atom, reflecting a significant destabilizing effect due to H interstitials, while the atomic volume expands to 19.40 Å$^3$. This volume expansion suggests that hydrogen introduces substantial lattice strain, altering interatomic distances and promoting local lattice distortions.

For the $Ti_5Mo_{95}$ alloy, the negative formation energy of -33.8 meV/atom indicates a thermodynamically stable phase, with minimal volume change after relaxation (−2.7%-2.7\%−2.7%), suggesting limited structural distortion. However, when hydrogen is introduced ($Ti_5Mo_{95}$+H), the formation energy sharply increases to 660.6 meV/atom, and the atomic volume per atom expands to 19.75 Å$^3$, indicating strong lattice distortion effects. The inclusion of oxygen impurities further exacerbates these effects, as seen in $Ti_5Mo_{95}$+H [O1+O2], where the formation energy rises to 715.4 meV/atom, and the atomic volume expands significantly to 25.35 Å$^3$ per atom. This highlights the combined impact of H interstitials and oxygen-induced lattice strain, which leads to substantial alterations in phase stability and mechanical behavior.

**Table A1.** Analysis of formation energy ($E_{form}$), local lattice distortion (LLD), and volume change in BCC MPEAs with hydrogen incorporation at energetically favorable octahedral sites. ALA6 (Ames Lab Alloy #6) is the alloy composition developed at Ames, $Mo_{72.3}W_{12.8}Ta_{10}Ti_{2.5}Zr_{2.5}$ [86,87].

| MPEAs | $E_{form}$ [meV/atom] | | %Change | V |
|---|---|---|---|---|
| | Ideal | Relaxed [LLD] | [Ideal-RLX]/Ideal | [Å$^3$ per atom] |
| **Mo** | 0 | 0 | 0 | 15.8 |
| **Mo + H [O]** | 624.5 | 115.3 | 81.5 | 19.4 |
| **$Ti_5Mo_{95}$** | -33.8 | -34.7 | -2.7 | 15.8 |
| **$Ti_5Mo_{95}$ + H [O]** | 660.6 | 81.4 | 87.7 | 19.8 |
| **$Ti_5Mo_{95}$ + H [O1+O2]** | 715.4 | 194.9 | 72.8 | 25.4 |
| **ALA6** | -14.9 | -23.4 | -57.0 | 16.1 |
| **ALA6 + H** | 623.5 | 82.0 | 86.8 | 19.6 |

We also analyzed a Mo-rich high-entropy composition, i.e., ALA6 ($Mo_{72.3}W_{12.8}Ta_{10}Ti_{2.5}Zr_{2.5}$ [86,87]), which consists of multiple elements, the formation energy remains negative at -14.9 meV/atom, signifying stability. The equilibrium atomic volume is 16.12 Å$^3$, larger than that of pure Mo. Upon hydrogen incorporation (ALA6 + H), however, the formation energy rises sharply to 623.5 meV/atom, demonstrating a destabilizing effect similar to Mo-based systems. The volume expansion is also pronounced, with the per-cell volume increasing from 1934.4 Å$^3$ to 2351.1 Å$^3$ and the per-atom volume reaching 19.59 Å$^3$, reinforcing the role of hydrogen in modifying the local atomic structure and increasing lattice strain. Overall, the results demonstrate that hydrogen significantly destabilizes these alloys by increasing formation energy and inducing local lattice distortions, with oxygen further amplifying these effects in Mo-based systems. The observed volumetric expansion in hydrogenated alloys indicates that interstitial hydrogen alters atomic spacing, affecting phase stability and mechanical properties. These insights are particularly relevant for understanding hydrogen embrittlement, phase transformations, and structural integrity in refractory MPEAs, which are critical for applications in hydrogen storage and high-temperature environments.

**Appendix 1.2 Relaxation and hydrogen effect on electronic structure**

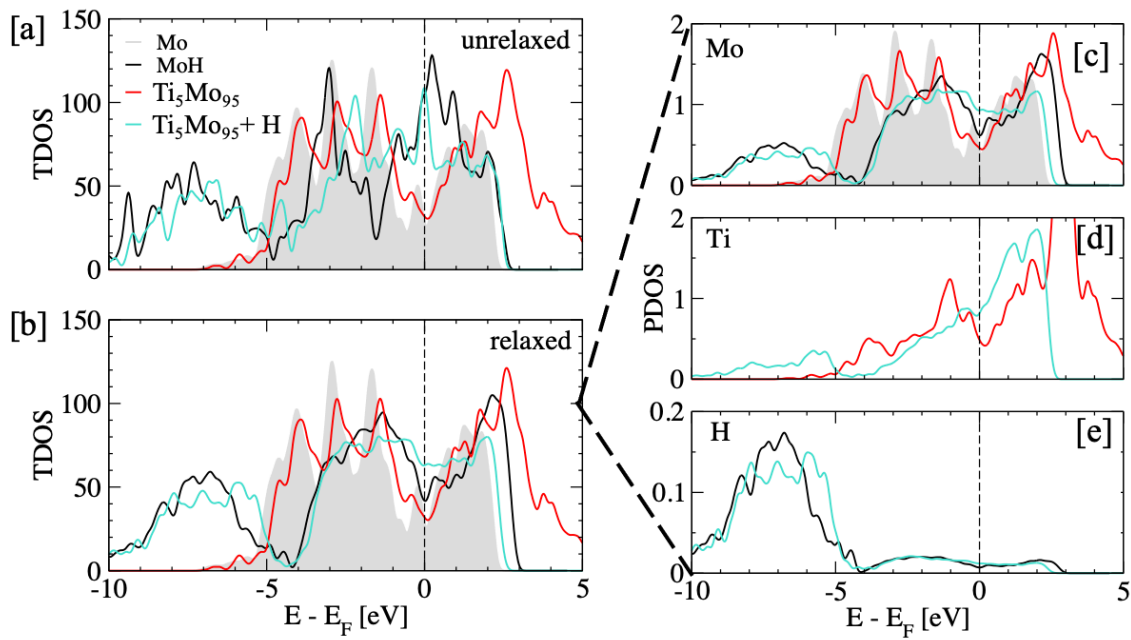

**Figure A1**. Total and projected density of states (DOS) for Mo, MoH, Ti$_5$Mo$_{95}$, and Ti$_5$Mo$_{95}$+H in (a) unrelaxed, and (b) relaxed structures, indicating significant electronic redistribution upon relaxation. (c-e) The PDOS for Mo, Ti, and H, respectively, highlighting the weak metal-hydrogen interactions in Ti$_5$Mo$_{95}$+H. The absence of H-states near the E$_{Fermi}$ confirms unfavourability of Ti$_5$Mo$_{95}$ for storage. The relaxation removed the sharp peak at the Fermi level in Mo-H that suggests electronic instability [**82**], which still indicates poor hydrogen storage capability.

In this section, we discussed the effect of relaxation on the electronic-structure of Mo, Mo-H, Ti-Mo, and Ti-Mo-H alloys. The presence of a sharp peak in the Mo-H TDOS at the E$_{Fermi}$ in **Fig. A1** shows electronic instability, making it difficult to maintain the crystal integrity upon hydrogen absorption. A high density of states at $E_{Fermi}$ indicates that the system is prone to electronic excitations, which can lead to structural distortions and phase instability. This instability results in weak metal-hydrogen interactions, reducing the material's ability to effectively store hydrogen. In the case of $Ti_5Mo_{95}$, the unrelaxed state shows a pronounced peak at $E_{Fermi}$, implying poor stability when hydrogen is introduced. Upon relaxation, the redistribution of states reduces the intensity of this peak, slightly improving stability. However, the fundamental electronic structure still lacks the necessary characteristics for effective hydrogen retention. The lack hydrogen states near the Fermi-level, weak bonding environment, and the presence of unstable electronic states make it challenging for the material to maintain its structural integrity under hydrogenation, further confirming that $Ti_5Mo_{95}$ is a poor hydrogen storage candidate.

**Appendix 1.3 Validation of weak hydrogen absorption in Mo-rich alloys**: The hydrogen absorption behavior of Ti and $Ti_5Mo_{95}$ alloys, showing temperature (T) and hydrogen content (wt%) as functions of time (t). Ti and $Ti_5Mo_{95}$ compositions were prepared by arc melting of pure elemental ingots on a water-cooled copper hearth, with five cycles of flipping and remelting the alloy ingot to achieve higher compositional uniformity. The hydrogen absorption and desorption tests in Ti and Mo-Ti were done by placing the samples into a pressure-composition isotherm (PCT) experiments (PCT) using Pro2000 system that allows for the control of pressure and temperature for the addition/removal of hydrogen to/from material systems. With this setup, after dynamic vacuum activation at 375°C, we expose samples to a high-pressure (~25 bar) hydrogen atmosphere and monitor the pressure changes during absorption/desorption at temperatures ranging from room temperature to 375°C. The amount of hydrogen absorbed was determine by monitoring the change in hydrogen pressure during measurement along with calibrated volumes and the ideal gas law.

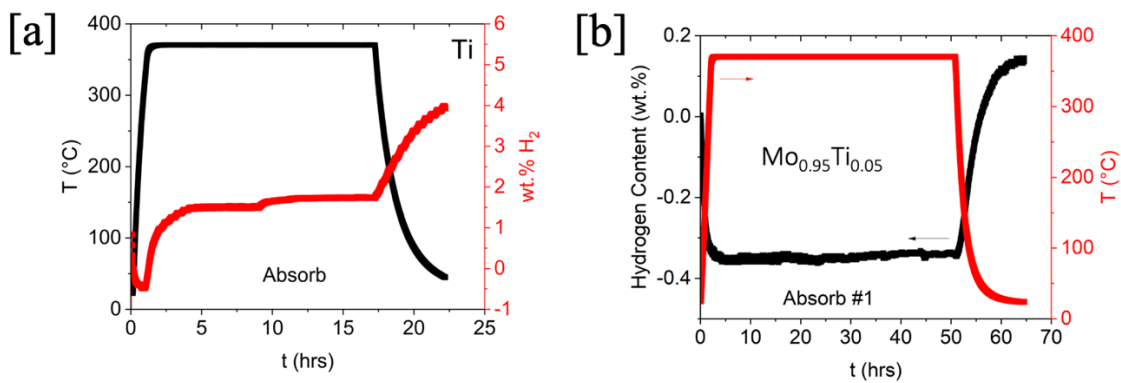

**Figure A2.** Hydrogen absorption behavior of (a) Ti and (b) $Ti_5Mo_{95}$ alloys. Ti shows rapid hydrogen uptake (~3.0 wt.%) while $Ti_5Mo_{95}$ absorbs less hydrogen (~0.2 wt.%) due to slower kinetics.

Our experiments on hydrogen absorption process for Ti (**Fig. A2a**) and Ti$_5$Mo$_{95}$ (**Fig. A2b**) also confirms our predictions that Mo-based alloys have slower absorption kinetics and lower capacity, likely due to the presence of Mo, which weakens hydrogen solubility (see **Fig. 8**). The desorption process also differs, with Ti releasing hydrogen more readily, whereas Ti$_5$Mo$_{95}$ shows a more extended desorption period. These results highlight the influence of alloying on hydrogen storage performance, with Ti being more suitable for high-capacity applications and Mo-containing alloys exhibiting lower absorption.